\documentclass[thmsa,a4paper,11pt]{article}%
\usepackage{amssymb}
\usepackage{amsmath}
\usepackage{longtable}%
\usepackage{amsfonts}%
\usepackage{graphicx}
\providecommand{\U}[1]{\protect\rule{.1in}{.1in}}
\begin{document}

\author{Dirk Veestraeten\thanks{Address correspondence to Dirk Veestraeten, Amsterdam
School of Economics, University of Amsterdam, Roetersstraat 11, 1018 WB
Amsterdam, The Netherlands, e-mail: d.j.m.veestraeten@uva.nl.}\\University of Amsterdam\\Amsterdam School of Economics\\Roetersstraat 11\\1018 WB Amsterdam\\The Netherlands\\\bigskip\bigskip}
\title{Some integral representations and limits for (products of) the parabolic
cylinder function \bigskip\bigskip}
\maketitle

\begin{abstract}
\noindent Veestraeten \cite{v15} recently derived inverse Laplace transforms
for Laplace transforms that contain products of two parabolic cylinder
functions by exploiting the link between the parabolic cylinder function and
the transition density and distribution functions of the Ornstein-Uhlenbeck
process. This paper first uses these results to derive new integral
representations for (products of two) parabolic cylinder functions. Second, as
the Brownian motion process with drift is a limiting case of the
Ornstein-Uhlenbeck process also limits can be calculated for the product of
gamma functions and (products of) parabolic cylinder functions. The central
results in both cases contain, in stylised form, $D_{v}\left(  x\right)
D_{v}\left(  y\right)  $ and $D_{v}\left(  x\right)  D_{v-1}\left(  y\right)
$ such that the recurrence relation of the parabolic cylinder function
straightforwardly allows to obtain integral representations and limits also
for countless other combinations in the orders such as $D_{v}\left(  x\right)
D_{v-3}\left(  y\right)  $ and $D_{v+1}\left(  x\right)  D_{v}\left(
y\right)  $.

\vspace{2cm}


\medskip


\newpage

\end{abstract}

\setlength{\baselineskip}{1.25\baselineskip}

\section{Introduction}

Veestraeten \cite{v15} recently derived inverse Laplace transforms for Laplace
transforms that include products of two parabolic cylinder functions of the
form $D_{v}\left(  x\right)  D_{v}\left(  y\right)  $ and $D_{v}\left(
x\right)  D_{v-1}\left(  y\right)  $ by exploiting the close link between the
parabolic cylinder function and the transition density and distribution
functions of the Ornstein-Uhlenbeck process. The recurrence relation for the
parabolic cylinder function then was used to obtain inverse transforms also
for products of parabolic cylinder functions such as $D_{v}\left(  x\right)
D_{v-3}\left(  y\right)  $ and $D_{v+1}\left(  x\right)  D_{v}\left(
y\right)  $.

This paper first uses these results to calculate integral representations for
products of two parabolic cylinder functions with differing arguments and that
have the same or different order. The result for the product in which the
parabolic cylinder functions share the same order corresponds with the
integral representations that were obtained in Malyshev \cite{m03} and Glasser
\cite{g15}. Our methodology also allows to derive multiple representations for
products that have the same combination of orders in the two parabolic
cylinder functions. Furthermore, for orders $-1$ and $-1/2$, the parabolic
cylinder function simplifies into the complementary error function,
erfc$\left(  x\right)  $, and the modified Bessel function of order $1/4$,
$K_{1/4}\left(  x\right)  $, respectively, such that also integral
representations can be obtained for erfc$\left(  x\right)  $erfc$\left(
y\right)  $, $K_{1/4}\left(  x\right)  K_{1/4}\left(  y\right)  $,
$K_{1/4}\left(  x\right)  $, $K_{1/4}\left(  x\right)  D_{-3/2}\left(
y\right)  $, etc.

Second, the Brownian motion process with drift can be obtained from the
Ornstein-Uhlenbeck process by letting the mean-reversion parameter $\beta$ in
the latter process approach $0$ (see Cox and Miller \cite{cm72}). This
property enables the derivation of limits for expressions that contain gamma
functions and (products of) parabolic cylinder functions in which the order in
the latter functions approaches negative infinity. These limits do not yield
asymptotic expansions for parabolic cylinder functions (see Olver \cite{o59}),
but evaluate into exponential functions that are independent of $\beta$ akin
to the results for other special functions that are obtained in Brychkov
\cite{b08}.

This paper illustrates the basic relations and presents some examples.
However, the recurrence relation of the parabolic cylinder function can be
used to extend results to various other orders in the (product of) parabolic
cylinder functions. The remainder of the paper is organised as follows.
Section $2$ lists the results of Veestraeten \cite{v15} that will be used
throughout the paper and subsequently derives integral representations for
(products of two) parabolic cylinder functions. Section $3$ discusses the
relation between the (Laplace transforms of the) Ornstein-Uhlenbeck process
and the Brownian motion process and obtains several limits that include gamma
functions and (products of) parabolic cylinder functions.

\bigskip

\section{Integral representations for (products of) parabolic cylinder
functions and related special functions}

The below integral representations and limits are calculated on the basis of
six inverse Laplace transforms that were derived in Veestraeten \cite{v15}.
For ease of reference, these results are reproduced in Table $1$.%
\[
\fbox{Table 1: around here.}%
\]
\noindent Using $v=-\tfrac{s+c}{\beta}$ in the first entry of Table $1$ gives
the following integral representation for the product of two parabolic
cylinder functions of the same order but with differing arguments%
\begin{align}
& D_{v}\left(  x\right)  D_{v}\left(  y\right)  =\dfrac{\exp\left(
\tfrac{y^{2}-x^{2}}{4}\right)  }{\Gamma\left(  -v\right)  }%
%TCIMACRO{\dint \limits_{0}^{+\infty}}%
%BeginExpansion
{\displaystyle\int\limits_{0}^{+\infty}}
%EndExpansion
\dfrac{\beta\exp\left[  \beta vt\right]  }{\sqrt{1-\exp\left(  -2\beta
t\right)  }}\exp\left(  -\dfrac{\left(  y+x\exp\left(  -\beta t\right)
\right)  ^{2}}{2\left(  1-\exp\left(  -2\beta t\right)  \right)  }\right)
dt\nonumber\\
& \hspace{0.5cm}\left[  \operatorname{Re}v<0,\beta>0,x+y\geqslant0\right]
.\tag{2.1}\label{2.1}%
\end{align}
\noindent The substitution $u=1-\exp\left(  -2\beta t\right)  $ allows to
write the latter result more compactly%
\begin{align}
& D_{v}\left(  x\right)  D_{v}\left(  y\right)  =\dfrac{\exp\left(
\tfrac{y^{2}-x^{2}}{4}\right)  }{2\Gamma\left(  -v\right)  }%
%TCIMACRO{\dint \limits_{0}^{1}}%
%BeginExpansion
{\displaystyle\int\limits_{0}^{1}}
%EndExpansion
\dfrac{1}{\left(  1-u\right)  ^{1+\frac{v}{2}}\sqrt{u}}\exp\left(
-\dfrac{\left(  y+x\sqrt{1-u}\right)  ^{2}}{2u}\right)  du\nonumber\\
& \hspace{0.5cm}\left[  \operatorname{Re}v<0,x+y\geqslant0\right]
,\tag{2.2}\label{2.2}%
\end{align}
\noindent which, together with the other central integral representations, is
listed in Table $2$.%
\[
\fbox{Table 2: around here.}%
\]
\noindent Glasser \cite{g15} recently derived the following integral
representation for $D_{-v}\left(  x\right)  D_{-v}\left(  -y\right)  $ under
$\operatorname{Re}v>0$ and $x>y>0$%
\begin{align*}
& D_{-v}\left(  x\right)  D_{-v}\left(  -y\right)  =\\
& \hspace{1cm}\dfrac{\exp\left(  -\tfrac{1}{4}\left(  x^{2}+y^{2}\right)
\right)  }{2\Gamma\left(  v\right)  }%
%TCIMACRO{\dint \limits_{0}^{\infty}}%
%BeginExpansion
{\displaystyle\int\limits_{0}^{\infty}}
%EndExpansion
t^{v/2-1}\left(  t+1\right)  ^{-\left(  v+1\right)  /2}\exp\left(  -\tfrac
{1}{2}\left(  x^{2}+y^{2}\right)  t+xy\sqrt{t\left(  t+1\right)  }\right)  dt.
\end{align*}
\noindent Using the substitutions $u=\tfrac{1}{q}$ and subsequently $q=t+1$
within the integral (\ref{2.2}) reveals that both expressions are identical.
Note that the requirement in Glasser \cite{g15} of the arguments having
opposite sign is too restrictive.

Applying the steps that were used for the integral representation (\ref{2.2})
to the second entry in Table $1$ produces%
\begin{align}
& D_{v}\left(  x\right)  D_{v-1}\left(  y\right)  =\dfrac{\sqrt{\pi}%
\exp\left(  \tfrac{y^{2}-x^{2}}{4}\right)  }{2^{\frac{3}{2}}\Gamma\left(
-v\right)  }%
%TCIMACRO{\dint \limits_{0}^{1}}%
%BeginExpansion
{\displaystyle\int\limits_{0}^{1}}
%EndExpansion
\dfrac{1}{\left(  1-u\right)  ^{1+\frac{v}{2}}}\text{erfc}\left(
\dfrac{y+x\sqrt{1-u}}{\sqrt{2u}}\right)  du\nonumber\\
& \hspace{0.5cm}\left[  \operatorname{Re}v<0,x+y\geqslant0\right]
,\tag{2.3}\label{2.3}%
\end{align}
\noindent where erfc$\left(  z\right)  $ is the complementary error function,
see Abramowitz and Stegun \cite{as72}. Setting $c=0$ in the third entry in
Table $1$ yields a further representation for $D_{v}\left(  x\right)
D_{v-1}\left(  y\right)  $ that instead relies on the exponential function%
\begin{align}
& D_{v}\left(  x\right)  D_{v-1}\left(  y\right)  =\nonumber\\
& \hspace{0.5cm}-\dfrac{\exp\left(  \tfrac{y^{2}-x^{2}}{4}\right)  }%
{2v\Gamma\left(  -v\right)  }%
%TCIMACRO{\dint \limits_{0}^{1}}%
%BeginExpansion
{\displaystyle\int\limits_{0}^{1}}
%EndExpansion
\dfrac{x+y\sqrt{1-u}}{\left(  1-u\right)  ^{\frac{1+v}{2}}u^{\frac{3}{2}}}%
\exp\left(  -\dfrac{\left(  y+x\sqrt{1-u}\right)  ^{2}}{2u}\right)  du-\left.
\dfrac{\sqrt{\pi}}{v\sqrt{2}\Gamma\left(  -v\right)  }\right\vert
_{x+y=0}\nonumber\\
& \hspace{1cm}\left[  \operatorname{Re}v<0,x+y\geqslant0\right]
.\tag{2.4}\label{2.4}%
\end{align}
\noindent Note that the term $\tfrac{\sqrt{\pi}}{v\sqrt{2}\Gamma\left(
-v\right)  }$ is to be subtracted when the sum of the arguments equals $0$ for
which the notation $\left.  \tfrac{\sqrt{\pi}}{v\sqrt{2}\Gamma\left(
-v\right)  }\right\vert _{x+y=0}$ is employed. This additional term originated
from the use of the differentiation property of the Laplace transform in
deriving the third entry in Table $1$, see Veestraeten \cite{v15} for more detail.

The inverse Laplace transforms in entries 4--6 in Table $1$ specify the
following three integral representations for $D_{v}\left(  x\right)
D_{v-2}\left(  y\right)  $%
\begin{align}
& D_{v}\left(  x\right)  D_{v-2}\left(  y\right)  =\dfrac{\exp\left(
\tfrac{y^{2}-x^{2}}{4}\right)  }{2\Gamma\left(  -v\right)  }%
%TCIMACRO{\dint \limits_{0}^{1}}%
%BeginExpansion
{\displaystyle\int\limits_{0}^{1}}
%EndExpansion
\dfrac{1}{\left(  1-u\right)  ^{1+\frac{v}{2}}}\left\{  \sqrt{u}\exp\left(
-\dfrac{\left(  y+x\sqrt{1-u}\right)  ^{2}}{2u}\right)  \right. \nonumber\\
& \hspace{0.5cm}\left.  -\sqrt{\tfrac{\pi}{2}}\left(  x\sqrt{1-u}+y\right)
\text{erfc}\left(  \dfrac{y+x\sqrt{1-u}}{\sqrt{2u}}\right)  \right\}
du\nonumber\\
& \hspace{1cm}\left[  \operatorname{Re}v<0,x+y\geqslant0\right]
,\tag{2.5}\label{2.5}%
\end{align}
\noindent and%
\begin{align}
& D_{v}\left(  x\right)  D_{v-2}\left(  y\right)  =-\dfrac{\exp\left(
\tfrac{y^{2}-x^{2}}{4}\right)  }{2v\Gamma\left(  -v\right)  }%
%TCIMACRO{\dint \limits_{0}^{1}}%
%BeginExpansion
{\displaystyle\int\limits_{0}^{1}}
%EndExpansion
\dfrac{1}{\left(  1-u\right)  ^{1+\frac{v}{2}}}\left\{  \dfrac{1-u}{\sqrt{u}%
}\exp\left(  -\dfrac{\left(  y+x\sqrt{1-u}\right)  ^{2}}{2u}\right)  \right.
\nonumber\\
& \hspace{0.5cm}\left.  +x\sqrt{\tfrac{\pi}{2}\left(  1-u\right)  }%
\text{erfc}\left(  \dfrac{y+x\sqrt{1-u}}{\sqrt{2u}}\right)  \right\}
du\nonumber\\
& \hspace{1cm}\left[  \operatorname{Re}v<0,x+y\geqslant0\right]
,\tag{2.6}\label{2.6}%
\end{align}
\noindent and%
\begin{align}
& D_{v}\left(  x\right)  D_{v-2}\left(  y\right)  =\dfrac{\exp\left(
\tfrac{y^{2}-x^{2}}{4}\right)  }{2\left(  1-v\right)  \Gamma\left(  -v\right)
}%
%TCIMACRO{\dint \limits_{0}^{1}}%
%BeginExpansion
{\displaystyle\int\limits_{0}^{1}}
%EndExpansion
\dfrac{1}{\left(  1-u\right)  ^{1+\frac{v}{2}}}\left\{  \dfrac{1}{\sqrt{u}%
}\exp\left(  -\dfrac{\left(  y+x\sqrt{1-u}\right)  ^{2}}{2u}\right)  \right.
\nonumber\\
& \hspace{0.5cm}\left.  -y\sqrt{\tfrac{\pi}{2}}\text{erfc}\left(
\dfrac{y+x\sqrt{1-u}}{\sqrt{2u}}\right)  \right\}  du\nonumber\\
& \hspace{1cm}\left[  \operatorname{Re}v<0,x+y\geqslant0\right]
.\tag{2.7}\label{2.7}%
\end{align}
\noindent The integral (\ref{2.5}) is a linear combination of integrals
(\ref{2.6}) and (\ref{2.7}) with weights $v$ and $\left(  1-v\right)  $,
respectively. The integral representations (\ref{2.6}) and (\ref{2.7}) mainly
differ in the definition of the weights of the exponential function and the
complementary error function with the weight of the latter function depending
on either $x$ or $y$, respectively. An additional, more compact integral
representation for $D_{v}\left(  x\right)  D_{v-2}\left(  y\right)  $ can be
obtained via the recurrence relation%
\begin{equation}
D_{v+1}\left(  z\right)  -zD_{v}\left(  z\right)  +vD_{v-1}\left(  z\right)
=0,\tag{2.8}\label{recurrence}%
\end{equation}
\noindent see Equation ($8.2.14$) in Erd\'{e}lyi et al. \cite{emoth253}.
Relation (\ref{recurrence}) yields%
\begin{align*}
& D_{v-2}\left(  y\right)  =\tfrac{y}{v-1}D_{v-1}\left(  y\right)  -\tfrac
{1}{v-1}D_{v}\left(  y\right)  ,\\
& D_{v}\left(  x\right)  D_{v-2}\left(  y\right)  =\tfrac{y}{v-1}D_{v}\left(
x\right)  D_{v-1}\left(  y\right)  -\tfrac{1}{v-1}D_{v}\left(  x\right)
D_{v}\left(  y\right)  .
\end{align*}
\noindent Plugging the integrals (\ref{2.4}) and (\ref{2.2}) into the latter
expression gives%
\begin{align}
& D_{v}\left(  x\right)  D_{v-2}\left(  y\right)  =\nonumber\\
& \hspace{0.5cm}\dfrac{\exp\left(  \tfrac{y^{2}-x^{2}}{4}\right)  }{2v\left(
1-v\right)  \Gamma\left(  -v\right)  }%
%TCIMACRO{\dint \limits_{0}^{1}}%
%BeginExpansion
{\displaystyle\int\limits_{0}^{1}}
%EndExpansion
\dfrac{y\sqrt{1-u}\left(  x+y\sqrt{1-u}\right)  +vu}{\left(  1-u\right)
^{1+\frac{v}{2}}u^{\frac{3}{2}}}\exp\left(  -\dfrac{\left(  y+x\sqrt
{1-u}\right)  ^{2}}{2u}\right)  du\nonumber\\
& \hspace{0.5cm}-\left.  \dfrac{y\sqrt{\pi}}{v\left(  v-1\right)  \sqrt
{2}\Gamma\left(  -v\right)  }\right\vert _{x+y=0}\nonumber\\
& \hspace{1cm}\left[  \operatorname{Re}v<0,x+y\geqslant0\right]
.\tag{2.9}\label{2.9}%
\end{align}
\noindent Likewise, the recurrence relation $D_{v}\left(  x\right)
D_{v+1}\left(  y\right)  =yD_{v}\left(  x\right)  D_{v}\left(  y\right)
-vD_{v}\left(  x\right)  D_{v-1}\left(  y\right)  $ together with integrals
(\ref{2.2}) and (\ref{2.4}) gives%
\begin{align}
& D_{v}\left(  x\right)  D_{v+1}\left(  y\right)  =\dfrac{\exp\left(
\tfrac{y^{2}-x^{2}}{4}\right)  }{2\Gamma\left(  -v\right)  }%
%TCIMACRO{\dint \limits_{0}^{1}}%
%BeginExpansion
{\displaystyle\int\limits_{0}^{1}}
%EndExpansion
\dfrac{yu+\sqrt{1-u}\left(  x+y\sqrt{1-u}\right)  }{\left(  1-u\right)
^{1+\frac{v}{2}}u^{\frac{3}{2}}}\exp\left(  -\dfrac{\left(  y+x\sqrt
{1-u}\right)  ^{2}}{2u}\right)  du\nonumber\\
& \hspace{0.5cm}+\left.  \dfrac{\sqrt{\pi}}{\sqrt{2}\Gamma\left(  -v\right)
}\right\vert _{x+y=0}\nonumber\\
& \hspace{1cm}\left[  \operatorname{Re}v<0,x+y\geqslant0\right]
.\tag{2.10}\label{2.10}%
\end{align}
\noindent Repeating the above steps also produces (multiple) integral
representations for products of parabolic cylinder functions that have more
distant orders such as $D_{v}\left(  x\right)  D_{v+5}\left(  y\right)  $ and
$D_{v}\left(  x\right)  D_{v-5}\left(  y\right)  $.

The above integral representations are valid for $x+y\geqslant0$ such that
they can easily be modified to allow for specific values in the arguments. For
instance, applying $\beta=1$ and $y=-x$ to Equation (\ref{2.1}) gives%
\begin{align*}
& D_{v}\left(  x\right)  D_{v}\left(  -x\right)  =\dfrac{1}{\Gamma\left(
-v\right)  }%
%TCIMACRO{\dint \limits_{0}^{+\infty}}%
%BeginExpansion
{\displaystyle\int\limits_{0}^{+\infty}}
%EndExpansion
\dfrac{\exp\left[  vt\right]  }{\sqrt{1-\exp\left(  -2t\right)  }}\exp\left(
-\dfrac{x^{2}\left(  1-\exp\left(  -t\right)  \right)  ^{2}}{2\left(
1-\exp\left(  -2t\right)  \right)  }\right)  dt\\
& \hspace{0.5cm}\left[  \operatorname{Re}v<0\right]  ,
\end{align*}
\noindent which now holds for all real arguments. The latter integral, given
the identity $\tanh\dfrac{t}{2}=\left(  \cosh t-1\right)  /\sinh t$, can be
rewritten as%
\begin{align*}
& D_{v}\left(  x\right)  D_{v}\left(  -x\right)  =2^{-1/2}\Gamma^{-1}\left(
-v\right)
%TCIMACRO{\dint \limits_{0}^{+\infty}}%
%BeginExpansion
{\displaystyle\int\limits_{0}^{+\infty}}
%EndExpansion
\exp\left(  \left(  v+\dfrac{1}{2}\right)  t-\dfrac{x^{2}}{2}\tanh\dfrac{t}%
{2}\right)  \dfrac{dt}{\sqrt{\sinh t}}\\
& \hspace{0.5cm}\left[  \operatorname{Re}v<0\right]  ,
\end{align*}
\noindent which is the result in Equation (6) of Malyshev \cite{m03}.

The above expressions also generate (multiple) representations for single
parabolic cylinder functions. This is illustrated using Equation (\ref{2.3})
for $y=0$ together with the identity%
\begin{equation}
D_{v}\left(  0\right)  =\dfrac{2^{v/2}\sqrt{\pi}}{\Gamma\left(  \dfrac{1-v}%
{2}\right)  },\tag{2.11}\label{pcf0}%
\end{equation}
\noindent see Magnus, Oberhettinger and Soni \cite{mos66}, where
$\Gamma\left(  v\right)  $ denotes the gamma function. Further simplifications
via the recurrence and the duplication formulas for the gamma function, i.e.
$\Gamma\left(  v+1\right)  =v\Gamma\left(  v\right)  $ and $\Gamma\left(
2v\right)  =\dfrac{1}{\sqrt{2\pi}}2^{2v-\frac{1}{2}}\Gamma\left(  v\right)
\Gamma\left(  v+\frac{1}{2}\right)  $ in Equations ($6.1.15$) and ($6.1.18$)
in Abramowitz and Stegun \cite{as72}, then give%
\begin{align*}
& D_{v}\left(  x\right)  =-\dfrac{v\sqrt{\pi}2^{\tfrac{1}{2}v-1}\exp\left(
\tfrac{-x^{2}}{4}\right)  }{\Gamma\left(  \tfrac{1}{2}-\tfrac{1}{2}v\right)  }%
%TCIMACRO{\dint \limits_{0}^{1}}%
%BeginExpansion
{\displaystyle\int\limits_{0}^{1}}
%EndExpansion
\dfrac{1}{\left(  1-u\right)  ^{1+\frac{v}{2}}}\text{erfc}\left(
\dfrac{x\sqrt{1-u}}{\sqrt{2u}}\right)  du\\
& \hspace{0.5cm}\left[  \operatorname{Re}v<0,x\geqslant0\right]  .
\end{align*}
\noindent The latter integral is expressed in terms of the complementary error
function and has not been documented in, for instance, Erd\'{e}lyi et al.
\cite{emoth253} and Gradshteyn and Ryzhik \cite{gr07}.

Choosing specific values for the order creates integral representations for
(products of) a number of other special functions. For instance, the parabolic
cylinder function simplifies into the complementary error function for $v=-1$
via%
\[
D_{-1}\left(  z\right)  =\sqrt{\dfrac{\pi}{2}}\exp\left(  \dfrac{z^{2}}%
{4}\right)  \text{erfc}\left(  \dfrac{z}{\sqrt{2}}\right)  ,
\]
\noindent see Equation ($9.254.1$) in Gradshteyn and Ryzhik \cite{gr07}. Using
$\beta=s+c$ within the first entry of Table $1$, applying the substitution
$u=1-\exp\left(  -2\beta t\right)  $ and rescaling $x$ and $y$ gives the
following compact integral representation for the product of two complementary
error functions
\begin{align*}
& \text{erfc}\left(  x\right)  \text{erfc}\left(  y\right)  =\dfrac
{\exp\left(  -x^{2}\right)  }{\pi}%
%TCIMACRO{\dint \limits_{0}^{1}}%
%BeginExpansion
{\displaystyle\int\limits_{0}^{1}}
%EndExpansion
\dfrac{1}{\sqrt{u\left(  1-u\right)  }}\exp\left(  -\dfrac{\left(
y+x\sqrt{1-u}\right)  ^{2}}{u}\right)  du\\
& \hspace{0.5cm}\left[  x+y\geqslant0\right]  ,
\end{align*}
\noindent which has not been documented in, for instance, the overview of Ng
and Geller \cite{ng69}.

The parabolic cylinder function with order $v=-\tfrac{1}{2}$ is related to the
modified Bessel function with order $\tfrac{1}{4}$ via%
\[
D_{-\tfrac{1}{2}}\left(  z\right)  =\dfrac{\sqrt{z}}{\sqrt{2\pi}}K_{\tfrac
{1}{4}}\left(  \dfrac{1}{4}z^{2}\right)  ,
\]
\noindent see Magnus, Oberhettinger and Soni \cite{mos66}. Setting $y=0$ in
the first entry of Table $1$ and using $\tfrac{1}{2}\beta=s+c$ yields%
\begin{align*}
& K_{1/4}\left(  x\right)  =\dfrac{\sqrt{\pi}}{\left(  2x\right)  ^{\frac
{1}{4}}\Gamma\left[  \frac{1}{4}\right]  }%
%TCIMACRO{\dint \limits_{0}^{1}}%
%BeginExpansion
{\displaystyle\int\limits_{0}^{1}}
%EndExpansion
\dfrac{1}{\left(  1-u\right)  ^{\frac{3}{4}}\sqrt{u}}\exp\left(
-\dfrac{x\left(  2-u\right)  }{u}\right)  du\\
& \hspace{0.5cm}\left[  x>0\right]  ,
\end{align*}
\noindent and an alternative representation in terms of the complementary
error function can be obtained from entry $2$ in Table $1$.

The above results also generate integral representations for combinations of
the aforementioned three special functions. For instance, entry $2$ in Table
$1$ with $\tfrac{1}{2}\beta=s+c$ gives%
\begin{align*}
& K_{\tfrac{1}{4}}\left(  x\right)  D_{-\tfrac{3}{2}}\left(  y\right)
=\dfrac{\sqrt{\pi}\exp\left(  \dfrac{y^{2}-4x}{4}\right)  }{2^{\frac{3}{2}%
}x^{\frac{1}{4}}}%
%TCIMACRO{\dint \limits_{0}^{1}}%
%BeginExpansion
{\displaystyle\int\limits_{0}^{1}}
%EndExpansion
\dfrac{1}{\left(  1-u\right)  ^{\frac{3}{4}}}\text{erfc}\left(  \dfrac
{y+2\sqrt{x\left(  1-u\right)  }}{\sqrt{2u}}\right)  du\\
& \hspace{0.5cm}\left[  x>0,2\sqrt{x}+y\geqslant0\right]  .
\end{align*}

\section{Limits containing (products of) parabolic cylinder functions}

The Brownian motion process with drift is a limiting case of the
Ornstein-Uhlenbeck process (see Cox and Miller \cite{cm72}) and this property
then also applies to the Laplace transforms for the transition density and
distribution functions of both processes. This section first lists the Laplace
transforms for the Ornstein-Uhlenbeck process that were derived in Veestraeten
\cite{v15} and subsequently obtains the transforms for the Brownian motion
process with drift.

Let $W=\{W_{t},$ $t\geqslant0\}$ be an Ornstein-Uhlenbeck process with initial
value $W_{0}=w_{0}$. The dynamics of this stochastic process are%
\begin{equation}
dW_{t}=\left(  \alpha-\beta W_{t}\right)  dt+\sigma dZ_{t}\text{,}%
\tag{3.1}\label{ou}%
\end{equation}
\noindent with $\beta>0$ and $t>0$. The instantaneous variance is $\sigma^{2}%
$, with $\sigma>0$, and $dZ_{t}$ denotes the increment of a Wiener process.
The process (\ref{ou}) is mean-reverting since the positive value of $\beta$
ensures that the stochastic variable reverts to its long-term mean
$\tfrac{\alpha}{\beta}$.

The transition probability distribution function -- in short, the transition
distribution -- is defined as $P\left(  w,t|w_{0}\right)  =\Pr\left\{
W\left(  t\right)  \leqslant w|W\left(  0\right)  =w_{0}\right\}  $ and the
transition probability density function -- in short, the transition density --
is given by $p\left(  w,t|w_{0}\right)  =\tfrac{\partial}{\partial w}P\left(
w,t|w_{0}\right)  $. The Laplace transform of the transition density of the
Ornstein-Uhlenbeck process, $\overline{p}\left(  w,s|w_{0}\right)  $, was
derived in Equation ($3.3$) in Veestraeten \cite{v15} as%
\begin{equation}
\overline{p}\left(  w,s|w_{0}\right)  =\left\{
\begin{array}
[c]{l}%
\dfrac{\Gamma\left(  s/\beta\right)  }{\sigma\sqrt{\pi\beta}}\exp\left(
\dfrac{\left(  \beta w_{0}-\alpha\right)  ^{2}}{2\sigma^{2}\beta}%
-\dfrac{\left(  \beta w-\alpha\right)  ^{2}}{2\sigma^{2}\beta}\right) \\
\ \hspace{0.5cm}\times D_{-s/\beta}\left(  \dfrac{\sqrt{2}\left(  \beta
w-\alpha\right)  }{\sigma\sqrt{\beta}}\right)  D_{-s/\beta}\left(
-\dfrac{\sqrt{2}\left(  \beta w_{0}-\alpha\right)  }{\sigma\sqrt{\beta}%
}\right)  \text{ for}-\infty\leqslant w_{0}\leqslant w,\\
\dfrac{\Gamma\left(  s/\beta\right)  }{\sigma\sqrt{\pi\beta}}\exp\left(
\dfrac{\left(  \beta w_{0}-\alpha\right)  ^{2}}{2\sigma^{2}\beta}%
-\dfrac{\left(  \beta w-\alpha\right)  ^{2}}{2\sigma^{2}\beta}\right) \\
\ \hspace{0.5cm}\times D_{-s/\beta}\left(  -\dfrac{\sqrt{2}\left(  \beta
w-\alpha\right)  }{\sigma\sqrt{\beta}}\right)  D_{-s/\beta}\left(
\dfrac{\sqrt{2}\left(  \beta w_{0}-\alpha\right)  }{\sigma\sqrt{\beta}%
}\right)  \text{ for }w\leqslant w_{0}\leqslant+\infty,
\end{array}
\right. \tag{3.2}\label{lttransdens}%
\end{equation}
\noindent and the Laplace transform for the transition distribution,
$\overline{P}\left(  w_{1},s|w_{0}\right)  $ with $w_{1}\geqslant w_{0}$, was
specified in Equation ($3.5$) as%
\begin{align}
& \overline{P}\left(  w_{1},s|w_{0}\right)  =\dfrac{1}{s}-\dfrac{\Gamma\left(
s/\beta\right)  }{\beta\sqrt{2\pi}}\exp\left(  \dfrac{\left(  \beta
w_{0}-\alpha\right)  ^{2}}{2\sigma^{2}\beta}-\dfrac{\left(  \beta w_{1}%
-\alpha\right)  ^{2}}{2\sigma^{2}\beta}\right) \tag{3.3}\label{lttransdist}\\
& \hspace{2cm}\times D_{-s/\beta}\left(  -\dfrac{\sqrt{2}\left(  \beta
w_{0}-\alpha\right)  }{\sigma\sqrt{\beta}}\right)  D_{-1-s/\beta}\left(
\dfrac{\sqrt{2}\left(  \beta w_{1}-\alpha\right)  }{\sigma\sqrt{\beta}%
}\right)  \text{ for }w_{1}\geqslant w_{0}.\nonumber
\end{align}
\noindent The dynamic process for Brownian motion with drift coefficient
$\alpha$ is given by%
\[
dW_{t}=\alpha dt+\sigma dZ_{t}\text{,}%
\]
\noindent with $t>0$ and thus can be seen as the limit of the
Ornstein-Uhlenbeck process (\ref{ou}) for $\beta$ approaching $0$. The
transition density of the Brownian motion, $p^{BM}\left(  w,t|w_{0}\right)  $,
satisfies the Kolmogorov forward equation (see Cox and Miller \cite{cm72} and
Risken \cite{r89})%
\[
\frac{1}{2}\sigma^{2}\frac{\partial^{2}p^{BM}\left(  w,t|w_{0}\right)
}{\partial w^{2}}-\alpha\hspace{0.05cm}\frac{\partial p^{BM}\left(
w,t|w_{0}\right)  }{\partial w}=\frac{\partial p^{BM}\left(  w,t|w_{0}\right)
}{\partial t}.
\]
\noindent The initial condition is%
\[
p^{BM}\left(  w,t|w_{0}\right)  =\delta\left(  w-w_{0}\right)  \qquad\text{for
}t=0,
\]
\noindent where $\delta\left(  \cdot\right)  $ is the Dirac delta function
that implies that all initial probability mass is located on the initial state.

The Laplace transform of the transition density, $\overline{p}^{BM}\left(
w,s|w_{0}\right)  $, is defined as%
\[
\overline{p}^{BM}\left(  w,s|w_{0}\right)  =%
%TCIMACRO{\dint \limits_{0}^{+\infty}}%
%BeginExpansion
{\displaystyle\int\limits_{0}^{+\infty}}
%EndExpansion
\exp\left(  -st\right)  p^{BM}\left(  w,t|w_{0}\right)  dt,
\]
\noindent where $\operatorname{Re}s>0$. The Kolmogorov forward equation then
simplifies into%
\[
\frac{1}{2}\sigma^{2}\dfrac{d^{2}\overline{p}^{BM}\left(  w,s|w_{0}\right)
}{dw^{2}}-\alpha\hspace{0.05cm}\frac{d\overline{p}^{BM}\left(  w,s|w_{0}%
\right)  }{dw}-s\hspace{0.05cm}\overline{p}^{BM}\left(  w,s|w_{0}\right)
=-\delta\left(  w-w_{0}\right)
\]
\noindent and its solution is%
\begin{equation}
\overline{p}^{BM}\left(  w,s|w_{0}\right)  =\left\{
\begin{array}
[c]{l}%
\dfrac{1}{\sqrt{\alpha^{2}+2s\sigma^{2}}}\exp\left(  \dfrac{-\alpha
+\sqrt{\alpha^{2}+2s\sigma^{2}}}{\sigma^{2}}\left(  w_{0}-w\right)  \right)
\text{ for}-\infty\leqslant w_{0}\leqslant w\text{ }\\
\dfrac{1}{\sqrt{\alpha^{2}+2s\sigma^{2}}}\exp\left(  \dfrac{-\alpha
-\sqrt{\alpha^{2}+2s\sigma^{2}}}{\sigma^{2}}\left(  w_{0}-w\right)  \right)
\text{ for }w\leqslant w_{0}\leqslant+\infty\text{.}%
\end{array}
\right. \tag{3.4}\label{lttransdensbm}%
\end{equation}
\noindent The transition distribution, $P^{BM}\left(  w_{1},t|w_{0}\right)  $,
specifies the cumulative probability mass that is located below $w_{1}$. Its
Laplace transform can be obtained by integrating the Laplace transform of the
transition density%
\[
\overline{P}^{BM}\left(  w_{1},s|w_{0}\right)  =%
%TCIMACRO{\dint \limits_{-\infty}^{w_{1}}}%
%BeginExpansion
{\displaystyle\int\limits_{-\infty}^{w_{1}}}
%EndExpansion
\overline{p}^{BM}\left(  w,s|w_{0}\right)  dw,
\]
\noindent see Veestraeten \cite{v15}. Then, $\overline{P}^{BM}\left(
w_{1},s|w_{0}\right)  $ is given by%
\begin{align}
& \overline{P}^{BM}\left(  w_{1},s|w_{0}\right)  =\dfrac{1}{s}-\dfrac
{\sigma^{2}}{\sqrt{\alpha^{2}+2s\sigma^{2}}\left(  \sqrt{\alpha^{2}%
+2s\sigma^{2}}-\alpha\right)  }\exp\left(  \dfrac{-\alpha+\sqrt{\alpha
^{2}+2s\sigma^{2}}}{\sigma^{2}}\left(  w_{0}-w_{1}\right)  \right) \nonumber\\
& \hspace{1cm}\text{for }w_{1}\geqslant x_{0}.\tag{3.5}\label{lttransdistbm}%
\end{align}
\noindent The Laplace transforms for the transition densities
(\ref{lttransdens}) and (\ref{lttransdensbm}) are related as follows%
\begin{align}
& \underset{\beta\rightarrow0}{\lim}\left[  \dfrac{\Gamma\left(
s/\beta\right)  }{\sigma\sqrt{\pi\beta}}D_{-s/\beta}\left(  \dfrac{\sqrt
{2}\left(  \beta w-\alpha\right)  }{\sigma\sqrt{\beta}}\right)  D_{-s/\beta
}\left(  -\dfrac{\sqrt{2}\left(  \beta w_{0}-\alpha\right)  }{\sigma
\sqrt{\beta}}\right)  \right] \nonumber\\
& \hspace{0.5cm}=\dfrac{1}{\sqrt{\alpha^{2}+2s\sigma^{2}}}\exp\left(
\dfrac{\sqrt{\alpha^{2}+2s\sigma^{2}}}{\sigma^{2}}\left(  w_{0}-w\right)
\right)  \text{\hspace{0.5cm}for}-\infty\leqslant w_{0}\leqslant
w.\tag{3.6}\label{3.6}%
\end{align}
\noindent Setting $\sigma$, $w$ and $w_{0}$ at $\sqrt{2}$,$\ x$ and $y$,
respectively, gives%
\begin{align}
& \underset{\beta\rightarrow0}{\lim}\left[  \dfrac{\Gamma\left(
s/\beta\right)  }{\sqrt{\beta}}D_{-s/\beta}\left(  x\sqrt{\beta}-\dfrac
{\alpha}{\sqrt{\beta}}\right)  D_{-s/\beta}\left(  y\sqrt{\beta}+\dfrac
{\alpha}{\sqrt{\beta}}\right)  \right]  =\nonumber\\
& \hspace{0.5cm}\dfrac{\sqrt{2\pi}}{\sqrt{\alpha^{2}+4s}}\exp\left(
-\dfrac{\sqrt{\alpha^{2}+4s}}{2}\left(  x+y\right)  \right) \nonumber\\
& \hspace{1cm}\left[  \operatorname{Re}s>0\right]  ,\tag{3.7}\label{3.7}%
\end{align}
\noindent which, together with the other basic relations that follow, is also
listed in Table $3$. Again, $x$ and $y$ are real but, in contrast to the above
integral representations, no condition is required as to their sum. Note that
the limit (\ref{3.7}) also arises when instead evaluating the Laplace
transforms for the domain $w\leqslant w_{0}\leqslant+\infty$.%
\[
\fbox{Table 3: around here.}%
\]
\noindent The orders of the parabolic cylinder functions in the limits in this
paper thus approach negative infinity with arguments that vanish $(\alpha=0)$
or that approach $+\infty$ or $-\infty$ depending on the sign of $\alpha$.

The Laplace transforms for the transition distributions (\ref{lttransdist})
and (\ref{lttransdistbm}), using the above notation, yield%
\begin{align}
& \underset{\beta\rightarrow0}{\lim}\left[  \dfrac{\Gamma\left(
s/\beta\right)  }{\beta}D_{-s/\beta}\left(  y\sqrt{\beta}+\dfrac{\alpha}%
{\sqrt{\beta}}\right)  D_{-1-s/\beta}\left(  x\sqrt{\beta}-\dfrac{\alpha
}{\sqrt{\beta}}\right)  \right]  =\nonumber\\
& \hspace{0.5cm}\dfrac{2\sqrt{2\pi}}{\sqrt{\alpha^{2}+4s}\left(  \sqrt
{\alpha^{2}+4s}-\alpha\right)  }\exp\left(  -\dfrac{\sqrt{\alpha^{2}+4s}}%
{2}\left(  x+y\right)  \right) \nonumber\\
& \hspace{1cm}\left[  \operatorname{Re}s>0\right]  .\tag{3.8}\label{3.8}%
\end{align}
\noindent The recurrence relation (\ref{recurrence}) allows to expand the
above results to other orders. For instance, using $v=-1-s/\beta$ in the
recurrence relation, applying it to the second parabolic cylinder function in
the product and keeping the order of the first parabolic cylinder function at
$-s/\beta$ gives%
\begin{align*}
& \underset{\beta\rightarrow0}{\lim}\left[  \dfrac{\left(  s+\beta\right)
\Gamma\left(  s/\beta\right)  }{\beta^{\frac{3}{2}}}D_{-s/\beta}\left(
y\sqrt{\beta}+\dfrac{\alpha}{\sqrt{\beta}}\right)  D_{-2-s/\beta}\left(
x\sqrt{\beta}-\dfrac{\alpha}{\sqrt{\beta}}\right)  \right]  =\\
& \hspace{1cm}\underset{\beta\rightarrow0}{\lim}\left[  \dfrac{-\left(
x\beta-\alpha\right)  \Gamma\left(  s/\beta\right)  }{\beta}D_{-s/\beta
}\left(  y\sqrt{\beta}+\dfrac{\alpha}{\sqrt{\beta}}\right)  D_{-1-s/\beta
}\left(  x\sqrt{\beta}-\dfrac{\alpha}{\sqrt{\beta}}\right)  \right] \\
& \hspace{1cm}+\underset{\beta\rightarrow0}{\lim}\left[  \dfrac{\Gamma\left(
s/\beta\right)  }{\sqrt{\beta}}D_{-s/\beta}\left(  y\sqrt{\beta}+\dfrac
{\alpha}{\sqrt{\beta}}\right)  D_{-s/\beta}\left(  x\sqrt{\beta}-\dfrac
{\alpha}{\sqrt{\beta}}\right)  \right]  .
\end{align*}
\noindent The first term on the right-hand side equals $\tfrac{2\alpha
\sqrt{2\pi}}{\sqrt{\alpha^{2}+4s}\left(  \sqrt{\alpha^{2}+4s}-\alpha\right)
}\exp\left(  -\tfrac{\sqrt{\alpha^{2}+4s}}{2}\left(  x+y\right)  \right)  $ as
can be inferred from Equation (\ref{3.8}) and the second right-hand side term
is given in Equation (\ref{3.7}) such that%
\begin{align}
& \underset{\beta\rightarrow0}{\lim}\left[  \dfrac{\left(  s+\beta\right)
\Gamma\left(  s/\beta\right)  }{\beta^{\frac{3}{2}}}D_{-s/\beta}\left(
y\sqrt{\beta}+\dfrac{\alpha}{\sqrt{\beta}}\right)  D_{-2-s/\beta}\left(
x\sqrt{\beta}-\dfrac{\alpha}{\sqrt{\beta}}\right)  \right]  =\nonumber\\
& \hspace{0.5cm}\dfrac{\sqrt{2\pi}\left(  \alpha+\sqrt{\alpha^{2}+4s}\right)
}{\sqrt{\alpha^{2}+4s}\left(  \sqrt{\alpha^{2}+4s}-\alpha\right)  }\exp\left(
-\dfrac{\sqrt{\alpha^{2}+4s}}{2}\left(  x+y\right)  \right) \nonumber\\
& \hspace{1cm}\left[  \operatorname{Re}s>0\right]  .\tag{3.9}\label{3.9}%
\end{align}
\noindent Limits for other orders can straightforwardly be obtained by
applying the recurrence relation in Equation (\ref{recurrence}) to the above limits.

The limiting case for a single parabolic cylinder function can directly be
evaluated via the limits (\ref{3.7})--(\ref{3.9}) and the identity
(\ref{pcf0}). However, this is not attractive as it requires $\alpha=0$ and
thus obscures the role of $\alpha$. Instead, the limiting case for a single
parabolic cylinder function within the limit (\ref{3.6}) is obtained by using
$w_{0}=\tfrac{\alpha}{\beta}$. The right-hand side then amounts to $\tfrac
{1}{\sqrt{\alpha^{2}+2s\sigma^{2}}}\exp\left(  \tfrac{\sqrt{\alpha
^{2}+2s\sigma^{2}}}{\sigma^{2}}\left(  \tfrac{\alpha}{\beta}-w\right)
\right)  $, which is $0$ for $\alpha<0$, $+\infty$ for $\alpha>0$ and
$\tfrac{1}{\sqrt{2s\sigma^{2}}}\exp\left(  -\tfrac{\sqrt{2s\sigma^{2}}}%
{\sigma^{2}}w\right)  $ for $\alpha=0$. Using $\sigma=\sqrt{2}$, $x=w$ and the
identity (\ref{pcf0}) gives%
\begin{align}
& \underset{\beta\rightarrow0}{\lim}\left[  \dfrac{2^{s/(2\beta)}\Gamma\left(
s/(2\beta)\right)  }{\sqrt{\beta}}D_{-s/\beta}\left(  x\sqrt{\beta}%
-\dfrac{\alpha}{\sqrt{\beta}}\right)  \right]  =\left\{
\begin{array}
[c]{ll}%
0\smallskip & \hspace{1cm}\alpha<0\\
\sqrt{\dfrac{2\pi}{s}}\exp\left(  -\sqrt{s}x\right)  \smallskip & \hspace
{1cm}\alpha=0\\
+\infty & \hspace{1cm}\alpha>0
\end{array}
\right.  \nonumber\\
& \hspace{1cm}\left[  \operatorname{Re}s>0\right]  .\tag{3.10}\label{3.10}%
\end{align}
\noindent The following two limits are obtained from the relation for the
transition distribution%
\begin{align}
& \underset{\beta\rightarrow0}{\lim}\left[  2^{s/(2\beta)}\Gamma\left(
\left(  s+\beta\right)  /\left(  2\beta\right)  \right)  D_{-s/\beta}\left(
x\sqrt{\beta}-\dfrac{\alpha}{\sqrt{\beta}}\right)  \right]  =\left\{
\begin{array}
[c]{ll}%
0\smallskip & \hspace{1cm}\alpha<0\\
\sqrt{\pi}\exp\left(  -\sqrt{s}x\right)  \smallskip & \hspace{1cm}\alpha=0\\
+\infty & \hspace{1cm}\alpha>0
\end{array}
\right.  \nonumber\\
& \hspace{1cm}\left[  \operatorname{Re}s>0\right]  ,\tag{3.11}\label{3.11}%
\end{align}
\noindent and%
\begin{align}
& \underset{\beta\rightarrow0}{\lim}\left[  \dfrac{2^{s/(2\beta)}\Gamma\left(
s/(2\beta)\right)  }{\beta}D_{-1-s/\beta}\left(  x\sqrt{\beta}-\dfrac{\alpha
}{\sqrt{\beta}}\right)  \right]  =\left\{
\begin{array}
[c]{ll}%
0\smallskip & \hspace{1cm}\alpha<0\\
\dfrac{\sqrt{2\pi}}{s}\exp\left(  -\sqrt{s}x\right)  \smallskip & \hspace
{1cm}\alpha=0\\
+\infty & \hspace{1cm}\alpha>0
\end{array}
\right.  \nonumber\\
& \hspace{1cm}\left[  \operatorname{Re}s>0\right]  .\tag{3.12}\label{3.12}%
\end{align}
\noindent Applying the recurrence relation (\ref{recurrence}) to the limits
(\ref{3.12}) and (\ref{3.10}) yields%
\begin{align}
& \underset{\beta\rightarrow0}{\lim}\left[  \dfrac{2^{s/(2\beta)}\left(
s+\beta\right)  \Gamma\left(  s/\left(  2\beta\right)  \right)  }{\beta
^{\frac{3}{2}}}D_{-2-s/\beta}\left(  x\sqrt{\beta}-\dfrac{\alpha}{\sqrt{\beta
}}\right)  \right]  =\left\{
\begin{array}
[c]{ll}%
0\smallskip & \hspace{1cm}\alpha<0\\
\sqrt{\dfrac{2\pi}{s}}\exp\left(  -\sqrt{s}x\right)  \smallskip & \hspace
{1cm}\alpha=0\\
+\infty & \hspace{1cm}\alpha>0
\end{array}
\right.  \nonumber\\
& \hspace{1cm}\left[  \operatorname{Re}s>0\right]  .\tag{3.13}\label{3.13}%
\end{align}
\noindent Note that setting the arguments in both parabolic cylinder functions
at $0$ within the above relations produces the following limit for the ratio
of gamma functions%
\begin{align}
& \underset{\beta\rightarrow0}{\lim}\left[  \dfrac{\Gamma\left(  s/\left(
2\beta\right)  \right)  }{\sqrt{\beta}\Gamma\left(  (s+\beta)/\left(
2\beta\right)  \right)  }\right]  =\sqrt{\dfrac{2}{s}}\nonumber\\
& \hspace{1cm}\left[  \operatorname{Re}s>0\right]  .\tag{3.14}\label{3.14}%
\end{align}
\noindent This result actually is a special case of the limit ($1.18.5$) in
Erd\'{e}lyi et al. \cite{emoth153}%
\[
\underset{\left\vert z\right\vert \rightarrow+\infty}{\lim}\left[  \exp\left(
-a\log\left(  z\right)  \right)  \Gamma\left(  z+a\right)  /\Gamma\left(
z\right)  \right]  =1,
\]
\noindent which in our notation can be written as $\underset{\beta
\rightarrow0}{\lim}\left[  \exp\left(  -a\log\left(  1/\beta\right)  \right)
\Gamma\left(  1/\beta+a\right)  /\Gamma\left(  1/\beta\right)  \right]  =1$.
Using $a=\frac{1}{2}$ in the latter limit and choosing $s=2$ in the limit
(\ref{3.14}) produces identical specifications.

The above results can also be used to evaluate limits for products of four
parabolic cylinder functions and various ratios. For instance, dividing the
limit (\ref{3.7}) by the limit (\ref{3.8}) gives%
\begin{align*}
& \underset{\beta\rightarrow0}{\lim}\left[  \left.  \sqrt{\beta}D_{-s/\beta
}\left(  x\sqrt{\beta}-\dfrac{\alpha}{\sqrt{\beta}}\right)  \right/
D_{-1-s/\beta}\left(  x\sqrt{\beta}-\dfrac{\alpha}{\sqrt{\beta}}\right)
\right]  =\dfrac{\sqrt{\alpha^{2}+4s}-\alpha}{2}\\
& \text{ }\hspace{1cm}\left[  \operatorname{Re}s>0\right]  ,
\end{align*}
\noindent which offers neat specialisations. For instance, using $x=0$ and
$\alpha=s=1$ gives a limit in which both the order as well as the argument
approach negative infinity%
\[
\underset{\beta\rightarrow0}{\lim}\left[  \left.  \sqrt{\beta}D_{-1/\beta
}\left(  -\dfrac{1}{\sqrt{\beta}}\right)  \right/  D_{-1-1/\beta}\left(
-\dfrac{1}{\sqrt{\beta}}\right)  \right]  =\dfrac{\sqrt{5}-1}{2}.
\]
\noindent Using $s=4$ and $x=1$ and letting the arguments approach $0$ via
$\alpha=0$ yields%
\[
\underset{\beta\rightarrow0}{\lim}\left[  \left.  \left(  \sqrt{\beta
}D_{-4/\beta}\left(  \sqrt{\beta}\right)  \right)  \right/  D_{-1-4/\beta
}\left(  \sqrt{\beta}\right)  \right]  =2.
\]

\newpage%

%TCIMACRO{\TeXButton{Table 1}{\begin{table}[h]\centering}}%
%BeginExpansion
\begin{table}[h]\centering
%EndExpansion
\caption{\ Inverse transforms of Laplace transforms that contain products of two
parabolic cylinder functions.*}\bigskip%

%TCIMACRO{\TeXButton{TeX field}{\small}}%
%BeginExpansion
\small
%EndExpansion%
\begin{tabular}
[c]{ll}\hline\hline
$\overline{f}\left(  s\right)  =%
%TCIMACRO{\dint \limits_{0}^{\infty}}%
%BeginExpansion
{\displaystyle\int\limits_{0}^{\infty}}
%EndExpansion
\exp\left(  -st\right)  \ f\left(  t\right)  \ dt\smallskip$ & $L^{-1}\left\{
\overline{f}\left(  s\right)  \right\}  =f\left(  t\right)  \smallskip
$\\\hline\hline
& \\
1. $\Gamma\left(  \dfrac{s+c}{\beta}\right)  D_{-\left(  s+c\right)  /\beta
}\left(  x\right)  D_{-\left(  s+c\right)  /\beta}\left(  y\right)  $ &
$\dfrac{\beta\exp\left(  -ct\right)  }{\sqrt{1-\exp\left(  -2\beta t\right)
}}\exp\left(  \dfrac{y^{2}-x^{2}}{4}\right)  \medskip$\\
& $\hspace{0.75cm}\times\exp\left(  -\dfrac{\left(  y+x\exp\left(  -\beta
t\right)  \right)  ^{2}}{2\left(  1-\exp\left(  -2\beta t\right)  \right)
}\right)  \medskip$\\
2. $\Gamma\left(  \dfrac{s+c}{\beta}\right)  D_{-\left(  s+c\right)  /\beta
}\left(  x\right)  D_{-1-\left(  s+c\right)  /\beta}\left(  y\right)  $ &
$\dfrac{\beta\sqrt{\pi}\exp\left(  -ct\right)  }{\sqrt{2}}\exp\left(
\dfrac{y^{2}-x^{2}}{4}\right)  $erfc$\left(  \dfrac{y+x\exp\left(  -\beta
t\right)  }{\sqrt{2\left(  1-\exp\left(  -2\beta t\right)  \right)  }}\right)
\medskip$\\
3. $s\Gamma\left(  \dfrac{s+c}{\beta}\right)  D_{-\left(  s+c\right)  /\beta
}\left(  x\right)  D_{-1-\left(  s+c\right)  /\beta}\left(  y\right)  $ &
$\exp\left(  \dfrac{y^{2}-x^{2}}{4}\right)  \left\{  \dfrac{\beta^{2}%
\exp\left(  -\left(  \beta+c\right)  t\right)  }{\left(  1-\exp\left(  -2\beta
t\right)  \right)  ^{\frac{3}{2}}}\right.  \medskip$\\
$\hspace{1cm}-\left.  \dfrac{\beta\sqrt{\pi}}{\sqrt{2}}\right\vert _{x+y=0}$ &
$\hspace{0.75cm}\times\left(  x+y\exp\left(  -\beta t\right)  \right)
\exp\left(  -\dfrac{\left(  y+x\exp\left(  -\beta t\right)  \right)  ^{2}%
}{2\left(  1-\exp\left(  -2\beta t\right)  \right)  }\right)  \medskip$\\
& $\left.  -\dfrac{c\beta\sqrt{\pi}\exp\left(  -ct\right)  }{\sqrt{2}%
}\text{erfc}\left(  \dfrac{y+x\exp\left(  -\beta t\right)  }{\sqrt{2\left(
1-\exp\left(  -2\beta t\right)  \right)  }}\right)  \right\}  \medskip$\\
4. $\Gamma\left(  \dfrac{s+c}{\beta}\right)  D_{-\left(  s+c\right)  /\beta
}\left(  x\right)  D_{-2-\left(  s+c\right)  /\beta}\left(  y\right)  $ &
$\exp\left(  \dfrac{y^{2}-x^{2}}{4}\right)  \left\{  \beta\sqrt{1-\exp\left(
-2\beta t\right)  }\exp\left(  -ct\right)  \right.  \medskip$\\
& $\hspace{0.75cm}\times\exp\left(  -\dfrac{\left(  y+x\exp\left(  -\beta
t\right)  \right)  ^{2}}{2\left(  1-\exp\left(  -2\beta t\right)  \right)
}\right)  \medskip$\\
& $-\dfrac{\left(  x\exp\left(  -\beta t\right)  +y\right)  \beta\sqrt{\pi
}\exp\left(  -ct\right)  }{\sqrt{2}}\medskip$\\
& $\hspace{0.75cm}\times\left.  \text{erfc}\left(  \dfrac{y+x\exp\left(
-\beta t\right)  }{\sqrt{2\left(  1-\exp\left(  -2\beta t\right)  \right)  }%
}\right)  \right\}  \medskip$\\
5. $\frac{s+c}{\beta}\Gamma\left(  \dfrac{s+c}{\beta}\right)  D_{-\left(
s+c\right)  /\beta}\left(  x\right)  D_{-2-\left(  s+c\right)  /\beta}\left(
y\right)  $ & $\exp\left(  \dfrac{y^{2}-x^{2}}{4}\right)  \left\{
\dfrac{\beta\exp\left(  -\left(  2\beta+c\right)  t\right)  }{\sqrt
{1-\exp\left(  -2\beta t\right)  }}\right.  \medskip$\\
& $\hspace{0.75cm}\times\exp\left(  -\dfrac{\left(  y+x\exp\left(  -\beta
t\right)  \right)  ^{2}}{2\left(  1-\exp\left(  -2\beta t\right)  \right)
}\right)  \medskip$\\
& $+\left.  \dfrac{x\beta\sqrt{\pi}\exp\left(  -\left(  \beta+c\right)
t\right)  }{\sqrt{2}}\text{erfc}\left(  \dfrac{y+x\exp\left(  -\beta t\right)
}{\sqrt{2\left(  1-\exp\left(  -2\beta t\right)  \right)  }}\right)  \right\}
\medskip$\\
6. $\frac{s+\beta+c}{\beta}\Gamma\left(  \dfrac{s+c}{\beta}\right)
D_{-\left(  s+c\right)  /\beta}\left(  x\right)  D_{-2-\left(  s+c\right)
/\beta}\left(  y\right)  $ & $\exp\left(  \dfrac{y^{2}-x^{2}}{4}\right)
\left\{  \dfrac{\beta\exp\left(  -ct\right)  }{\sqrt{1-\exp\left(  -2\beta
t\right)  }}\right.  \medskip$\\
& $\hspace{0.75cm}\times\exp\left(  -\dfrac{\left(  y+x\exp\left(  -\beta
t\right)  \right)  ^{2}}{2\left(  1-\exp\left(  -2\beta t\right)  \right)
}\right)  \medskip$\\
& $\left.  -\dfrac{y\beta\sqrt{\pi}\exp\left(  -ct\right)  }{\sqrt{2}%
}\text{erfc}\left(  \dfrac{y+x\exp\left(  -\beta t\right)  }{\sqrt{2\left(
1-\exp\left(  -2\beta t\right)  \right)  }}\right)  \right\}  \medskip
$\\\hline\hline
\multicolumn{2}{l}{for $\operatorname{Re}s>0,\beta>0,c\geqslant0,x+y\geqslant
0.$}%
\end{tabular}%
%TCIMACRO{\TeXButton{TeX field}{\normalsize}}%
%BeginExpansion
\normalsize
%EndExpansion%
%TCIMACRO{\TeXButton{E}{\end{table}}}%
%BeginExpansion
\end{table}%
%EndExpansion

\newpage%
%TCIMACRO{\TeXButton{Table 2}{\begin{table}[h]\centering}}%
%BeginExpansion
\begin{table}[h]\centering
%EndExpansion
\caption{\ Integral representations for products of two parabolic cylinder
functions.\label{key}}\bigskip%

%TCIMACRO{\TeXButton{TeX field}{\small}}%
%BeginExpansion
\small
%EndExpansion%
\begin{tabular}
[c]{c}\hline\hline
\\
\multicolumn{1}{l}{1. $D_{v}\left(  x\right)  D_{v}\left(  y\right)
=\dfrac{\exp\left(  \tfrac{y^{2}-x^{2}}{4}\right)  }{2\Gamma\left(  -v\right)
}%
%TCIMACRO{\dint \limits_{0}^{1}}%
%BeginExpansion
{\displaystyle\int\limits_{0}^{1}}
%EndExpansion
\dfrac{1}{\left(  1-u\right)  ^{1+\frac{v}{2}}\sqrt{u}}\exp\left(
-\dfrac{\left(  y+x\sqrt{1-u}\right)  ^{2}}{2u}\right)  du\medskip$}\\
\multicolumn{1}{l}{$\hspace{4cm}\left[  \operatorname{Re}v<0,x+y\geqslant
0\right]  \medskip$}\\
\multicolumn{1}{l}{2. $D_{v}\left(  x\right)  D_{v-1}\left(  y\right)
=\dfrac{\sqrt{\pi}\exp\left(  \tfrac{y^{2}-x^{2}}{4}\right)  }{2^{\frac{3}{2}%
}\Gamma\left(  -v\right)  }%
%TCIMACRO{\dint \limits_{0}^{1}}%
%BeginExpansion
{\displaystyle\int\limits_{0}^{1}}
%EndExpansion
\dfrac{1}{\left(  1-u\right)  ^{1+\frac{v}{2}}}\text{erfc}\left(
\dfrac{y+x\sqrt{1-u}}{\sqrt{2u}}\right)  du\medskip$}\\
\multicolumn{1}{l}{$\hspace{4cm}\left[  \operatorname{Re}v<0,x+y\geqslant
0\right]  \medskip$}\\
\multicolumn{1}{l}{3. $D_{v}\left(  x\right)  D_{v-1}\left(  y\right)
=-\dfrac{\exp\left(  \tfrac{y^{2}-x^{2}}{4}\right)  }{2v\Gamma\left(
-v\right)  }%
%TCIMACRO{\dint \limits_{0}^{1}}%
%BeginExpansion
{\displaystyle\int\limits_{0}^{1}}
%EndExpansion
\dfrac{x+y\sqrt{1-u}}{\left(  1-u\right)  ^{\frac{1+v}{2}}u^{\frac{3}{2}}}%
\exp\left(  -\dfrac{\left(  y+x\sqrt{1-u}\right)  ^{2}}{2u}\right)  du-\left.
\dfrac{\sqrt{\pi}}{v\sqrt{2}\Gamma\left(  -v\right)  }\right\vert
_{x+y=0}\medskip$}\\
\multicolumn{1}{l}{$\hspace{4cm}\left[  \operatorname{Re}v<0,x+y\geqslant
0\right]  \medskip$}\\
\multicolumn{1}{l}{4. $D_{v}\left(  x\right)  D_{v-2}\left(  y\right)
=\dfrac{\exp\left(  \tfrac{y^{2}-x^{2}}{4}\right)  }{2\Gamma\left(  -v\right)
}%
%TCIMACRO{\dint \limits_{0}^{1}}%
%BeginExpansion
{\displaystyle\int\limits_{0}^{1}}
%EndExpansion
\dfrac{1}{\left(  1-u\right)  ^{1+\frac{v}{2}}}\left\{  \sqrt{u}\exp\left(
-\dfrac{\left(  y+x\sqrt{1-u}\right)  ^{2}}{2u}\right)  \right.  $}\\
\multicolumn{1}{l}{$\hspace{3cm}\left.  -\sqrt{\tfrac{\pi}{2}}\left(
x\sqrt{1-u}+y\right)  \text{erfc}\left(  \dfrac{y+x\sqrt{1-u}}{\sqrt{2u}%
}\right)  \right\}  du\medskip$}\\
\multicolumn{1}{l}{$\hspace{4cm}\left[  \operatorname{Re}v<0,x+y\geqslant
0\right]  \medskip$}\\
\multicolumn{1}{l}{5. $D_{v}\left(  x\right)  D_{v-2}\left(  y\right)
=-\dfrac{\exp\left(  \tfrac{y^{2}-x^{2}}{4}\right)  }{2v\Gamma\left(
-v\right)  }%
%TCIMACRO{\dint \limits_{0}^{1}}%
%BeginExpansion
{\displaystyle\int\limits_{0}^{1}}
%EndExpansion
\dfrac{1}{\left(  1-u\right)  ^{1+\frac{v}{2}}}\left\{  \dfrac{1-u}{\sqrt{u}%
}\exp\left(  -\dfrac{\left(  y+x\sqrt{1-u}\right)  ^{2}}{2u}\right)  \right.
$}\\
\multicolumn{1}{l}{$\hspace{3cm}\left.  +x\sqrt{\tfrac{\pi}{2}\left(
1-u\right)  }\text{erfc}\left(  \dfrac{y+x\sqrt{1-u}}{\sqrt{2u}}\right)
\right\}  du\medskip$}\\
\multicolumn{1}{l}{$\hspace{4cm}\left[  \operatorname{Re}v<0,x+y\geqslant
0\right]  \medskip$}\\
\multicolumn{1}{l}{6. $D_{v}\left(  x\right)  D_{v-2}\left(  y\right)
=\dfrac{\exp\left(  \tfrac{y^{2}-x^{2}}{4}\right)  }{2\left(  1-v\right)
\Gamma\left(  -v\right)  }%
%TCIMACRO{\dint \limits_{0}^{1}}%
%BeginExpansion
{\displaystyle\int\limits_{0}^{1}}
%EndExpansion
\dfrac{1}{\left(  1-u\right)  ^{1+\frac{v}{2}}}\left\{  \dfrac{1}{\sqrt{u}%
}\exp\left(  -\dfrac{\left(  y+x\sqrt{1-u}\right)  ^{2}}{2u}\right)  \right.
$}\\
\multicolumn{1}{l}{$\hspace{3cm}\left.  -y\sqrt{\tfrac{\pi}{2}}\text{erfc}%
\left(  \dfrac{y+x\sqrt{1-u}}{\sqrt{2u}}\right)  \right\}  du\medskip$}\\
\multicolumn{1}{l}{$\hspace{4cm}\left[  \operatorname{Re}v<0,x+y\geqslant
0\right]  \medskip$}\\
\multicolumn{1}{l}{7. $D_{v}\left(  x\right)  D_{v-2}\left(  y\right)
=\dfrac{\exp\left(  \tfrac{y^{2}-x^{2}}{4}\right)  }{2v\left(  1-v\right)
\Gamma\left(  -v\right)  }%
%TCIMACRO{\dint \limits_{0}^{1}}%
%BeginExpansion
{\displaystyle\int\limits_{0}^{1}}
%EndExpansion
\dfrac{y\sqrt{1-u}\left(  x+y\sqrt{1-u}\right)  +vu}{\left(  1-u\right)
^{1+\frac{v}{2}}u^{\frac{3}{2}}}\exp\left(  -\dfrac{\left(  y+x\sqrt
{1-u}\right)  ^{2}}{2u}\right)  du$}\\
\multicolumn{1}{l}{$\hspace{3cm}-\left.  \dfrac{y\sqrt{\pi}}{v\left(
v-1\right)  \sqrt{2}\Gamma\left(  -v\right)  }\right\vert _{x+y=0}\medskip$}\\
\multicolumn{1}{l}{$\hspace{4cm}\left[  \operatorname{Re}v<0,x+y\geqslant
0\right]  \medskip$}\\
\multicolumn{1}{l}{8. $D_{v}\left(  x\right)  D_{v+1}\left(  y\right)
=\dfrac{\exp\left(  \tfrac{y^{2}-x^{2}}{4}\right)  }{2\Gamma\left(  -v\right)
}%
%TCIMACRO{\dint \limits_{0}^{1}}%
%BeginExpansion
{\displaystyle\int\limits_{0}^{1}}
%EndExpansion
\dfrac{yu+\sqrt{1-u}\left(  x+y\sqrt{1-u}\right)  }{\left(  1-u\right)
^{1+\frac{v}{2}}u^{\frac{3}{2}}}\exp\left(  -\dfrac{\left(  y+x\sqrt
{1-u}\right)  ^{2}}{2u}\right)  du+\left.  \dfrac{\sqrt{\pi}}{\sqrt{2}%
\Gamma\left(  -v\right)  }\right\vert _{x+y=0}\medskip$}\\
\multicolumn{1}{l}{$\hspace{4cm}\left[  \operatorname{Re}v<0,x+y\geqslant
0\right]  \medskip$}\\\hline\hline
\end{tabular}%
%TCIMACRO{\TeXButton{TeX field}{\normalsize}}%
%BeginExpansion
\normalsize
%EndExpansion%
%TCIMACRO{\TeXButton{E}{\end{table}}}%
%BeginExpansion
\end{table}%
%EndExpansion

\newpage%

%TCIMACRO{\TeXButton{Table 3}{\begin{table}[h]\centering}}%
%BeginExpansion
\begin{table}[h]\centering
%EndExpansion
\caption{\ Limits that contain gamma functions and (products of two)
parabolic cylinder functions.} \bigskip%

%TCIMACRO{\TeXButton{TeX field}{\small}}%
%BeginExpansion
\small
%EndExpansion%
\begin{tabular}
[c]{l}\hline\hline
\\
1. $\underset{\beta\rightarrow0}{\lim}\left[  \dfrac{\Gamma\left(
s/\beta\right)  }{\sqrt{\beta}}D_{-s/\beta}\left(  x\sqrt{\beta}-\dfrac
{\alpha}{\sqrt{\beta}}\right)  D_{-s/\beta}\left(  y\sqrt{\beta}+\dfrac
{\alpha}{\sqrt{\beta}}\right)  \right]  =\dfrac{\sqrt{2\pi}}{\sqrt{\alpha
^{2}+4s}}\exp\left(  -\dfrac{\sqrt{\alpha^{2}+4s}}{2}\left(  x+y\right)
\right)  \medskip$\\
$\hspace{4cm}\left[  \operatorname{Re}s>0\right]  \medskip$\\
2. $\underset{\beta\rightarrow0}{\lim}\left[  \dfrac{\Gamma\left(
s/\beta\right)  }{\beta}D_{-s/\beta}\left(  y\sqrt{\beta}+\dfrac{\alpha}%
{\sqrt{\beta}}\right)  D_{-1-s/\beta}\left(  x\sqrt{\beta}-\dfrac{\alpha
}{\sqrt{\beta}}\right)  \right]  =\medskip$\\
$\hspace{3cm}\dfrac{2\sqrt{2\pi}}{\sqrt{\alpha^{2}+4s}\left(  \sqrt{\alpha
^{2}+4s}-\alpha\right)  }\exp\left(  -\dfrac{\sqrt{\alpha^{2}+4s}}{2}\left(
x+y\right)  \right)  \medskip$\\
$\hspace{4cm}\left[  \operatorname{Re}s>0\right]  \medskip$\\
3. $\underset{\beta\rightarrow0}{\lim}\left[  \dfrac{\left(  s+\beta\right)
\Gamma\left(  s/\beta\right)  }{\beta^{\frac{3}{2}}}D_{-s/\beta}\left(
y\sqrt{\beta}+\dfrac{\alpha}{\sqrt{\beta}}\right)  D_{-2-s/\beta}\left(
x\sqrt{\beta}-\dfrac{\alpha}{\sqrt{\beta}}\right)  \right]  =\medskip$\\
$\hspace{3cm}\dfrac{\sqrt{2\pi}\left(  \alpha+\sqrt{\alpha^{2}+4s}\right)
}{\sqrt{\alpha^{2}+4s}\left(  \sqrt{\alpha^{2}+4s}-\alpha\right)  }\exp\left(
-\dfrac{\sqrt{\alpha^{2}+4s}}{2}\left(  x+y\right)  \right)  \medskip$\\
$\hspace{4cm}\left[  \operatorname{Re}s>0\right]  \medskip$\\
4. $\underset{\beta\rightarrow0}{\lim}\left[  \dfrac{2^{s/(2\beta)}%
\Gamma\left(  s/(2\beta)\right)  }{\sqrt{\beta}}D_{-s/\beta}\left(
x\sqrt{\beta}-\dfrac{\alpha}{\sqrt{\beta}}\right)  \right]  =\left\{
\begin{array}
[c]{ll}%
0\smallskip & \hspace{1cm}\alpha<0\\
\sqrt{\dfrac{2\pi}{s}}\exp\left(  -\sqrt{s}x\right)  \smallskip & \hspace
{1cm}\alpha=0\\
+\infty & \hspace{1cm}\alpha>0
\end{array}
\right.  \medskip$\\
$\hspace{4cm}\left[  \operatorname{Re}s>0\right]  \medskip$\\
5. $\underset{\beta\rightarrow0}{\lim}\left[  2^{s/(2\beta)}\Gamma\left(
\left(  s+\beta\right)  /\left(  2\beta\right)  \right)  D_{-s/\beta}\left(
x\sqrt{\beta}-\dfrac{\alpha}{\sqrt{\beta}}\right)  \right]  =\left\{
\begin{array}
[c]{ll}%
0\smallskip & \hspace{1cm}\alpha<0\\
\sqrt{\pi}\exp\left(  -\sqrt{s}x\right)  \smallskip & \hspace{1cm}\alpha=0\\
+\infty & \hspace{1cm}\alpha>0
\end{array}
\right.  \medskip$\\
$\hspace{4cm}\left[  \operatorname{Re}s>0\right]  \medskip$\\
6. $\underset{\beta\rightarrow0}{\lim}\left[  \dfrac{2^{s/(2\beta)}%
\Gamma\left(  s/(2\beta)\right)  }{\beta}D_{-1-s/\beta}\left(  x\sqrt{\beta
}-\dfrac{\alpha}{\sqrt{\beta}}\right)  \right]  =\left\{
\begin{array}
[c]{ll}%
0\smallskip & \hspace{1cm}\alpha<0\\
\dfrac{\sqrt{2\pi}}{s}\exp\left(  -\sqrt{s}x\right)  \smallskip & \hspace
{1cm}\alpha=0\\
+\infty & \hspace{1cm}\alpha>0
\end{array}
\right.  \medskip$\\
$\hspace{4cm}\left[  \operatorname{Re}s>0\right]  \medskip$\\
7. $\underset{\beta\rightarrow0}{\lim}\left[  \dfrac{2^{s/(2\beta)}\left(
s+\beta\right)  \Gamma\left(  s/\left(  2\beta\right)  \right)  }{\beta
^{\frac{3}{2}}}D_{-2-s/\beta}\left(  x\sqrt{\beta}-\dfrac{\alpha}{\sqrt{\beta
}}\right)  \right]  =\left\{
\begin{array}
[c]{ll}%
0\smallskip & \hspace{1cm}\alpha<0\\
\sqrt{\dfrac{2\pi}{s}}\exp\left(  -\sqrt{s}x\right)  \smallskip & \hspace
{1cm}\alpha=0\\
+\infty & \hspace{1cm}\alpha>0
\end{array}
\right.  \medskip$\\
$\hspace{4cm}\left[  \operatorname{Re}s>0\right]  \medskip$\\
8. $\underset{\beta\rightarrow0}{\lim}\left[  \dfrac{\Gamma\left(  s/\left(
2\beta\right)  \right)  }{\sqrt{\beta}\Gamma\left(  (s+\beta)/\left(
2\beta\right)  \right)  }\right]  =\sqrt{\dfrac{2}{s}}\medskip$\\
$\hspace{4cm}\left[  \operatorname{Re}s>0\right]  \medskip$\\\hline\hline
\end{tabular}%
%TCIMACRO{\TeXButton{TeX field}{\normalsize}}%
%BeginExpansion
\normalsize
%EndExpansion%
%TCIMACRO{\TeXButton{E}{\end{table}}}%
%BeginExpansion
\end{table}%
%EndExpansion

\end{document}